\documentclass{article}
\usepackage{amsmath}
\usepackage{url}
\usepackage{hyperref}
\usepackage{eufrak}
\usepackage{amsfonts}
\usepackage{amssymb}
\usepackage{amsthm}
\usepackage{color}

\usepackage{tikz}
\usepackage{tikzsymbols}
\usepackage[utf8]{inputenc} 
\usepackage{times}
\usepackage{mathtools}
\usepackage{footnote}
\makesavenoteenv{table}
\usetikzlibrary{arrows,shapes}
\usetikzlibrary{matrix}
\usetikzlibrary{shadows}
\usetikzlibrary{positioning}

\tikzset{
    >=stealth',
    pil/.style={
           ->,
           thick,
           shorten <=2pt,
           shorten >=2pt,}
}

\begin{document}

\title{\bf Approaches to spherically symmetric solutions in f(T) gravity}

\author{Alexey Golovnev${}^{1}$,  Mar\'ia-Jos\'e Guzm\'an${}^{2,3}$\\
{\small ${}^{1}${\it Centre for Theoretical Physics, British University in Egypt,}}\\ 
{\small\it 11837 El Sherouk City, Cairo, Egypt,}\\
{\small agolovnev@yandex.ru}\\
{\small ${}^{2}${\it Departamento de F\'isica y Astronom\'ia, Facultad de Ciencias, Universidad de La Serena,}}\\
{\small\it Av. Juan Cisternas 1200, 1720236 La Serena, Chile,}\\
{\small ${}^{3}${\it Laboratory of Theoretical Physics, Institute of Physics, University of Tartu,}}\\ 
{\small {\it W. Ostwaldi 1, 50411 Tartu, Estonia}}\\
{\small maria.j.guzman.m@gmail.com}
}
\date{}

\maketitle

\begin{abstract}

We study properties of static spherically symmetric solutions in $f(\mathbb T)$ gravity. Based on our previous work on generalising Bianchi identities for this kind of theories, we show how this search of solutions can be reduced to the study of two relatively simple equations. One of them does not depend on the function $f$ and therefore describes the properties of such solutions in any $f(\mathbb T)$ theory. Another equation is the radial one and, if a possible solution is chosen, it allows to find out which function $f$ is suitable for it. We use these equations to find exact and perturbative solutions for arbitrary and specific choices of $f$.

\end{abstract}

\section{Introduction}

Theories of modified teleparallel gravity are being actively studied now with plenty of different motivations, from solving the $H_0$ tension in cosmology to getting new angles at quantisation of gravitational interactions. This is an insurgent but hot topic in theoretical gravitational physics, which current development is dramatically speeding up \cite{Golovnev:2018red,Krssak:2018ywd}. In particular, the most used modified teleparallel models, the $f(\mathbb T)$ ones \cite{Ferraro:2006jd,Bengochea:2008gz,Ferraro:2008ey}, send us mixed messages about their perspectives and viability \cite{Ferraro:2018tpu,Blagojevic:2020dyq,Golovnev:2020nln,Golovnev:2020zpv}.

On one hand, construction of cosmological solutions in $f(\mathbb T)$ gravity has been utilised for model building, and quite successful when naively trusting the linear cosmological perturbation analysis \cite{Hashim:2020sez}. Partially, this success can be understood as a result of the number of dynamical degrees of freedom in linear perturbations being the same as in GR \cite{Golovnev:2018wbh}. It allows to easily compare the predictions and in particular to see non-vanishing gravitational slip as a possible trace of (modified) teleparallel nature of gravity.

On the other hand, non-perturbative Hamiltonian analyses show that there are extra modes, between one and three. Indicating the problem of strong coupling, it puts the $f(\mathbb T)$ cosmological models at serious doubt \cite{Golovnev:2020zpv}. Even though different analyses partially contradict each other \cite{Ferraro:2018tpu,Blagojevic:2020dyq}, existence of some new dynamical sector is firmly established, either by studying higher order perturbations around Minkowski spacetime, or at the level of linear perturbations around non-trivial tetrad representation of Minkowski which also presents shockingly puzzling behaviour \cite{Golovnev:2020nln, Golovnev:2020zpv}.

Given all these ideas being pursued, and while these models are increasingly widely used for cosmology, with relative success, we also need to make progress in understanding theoretical foundations as well as to aim at observational tests in different regimes. One of the most promising fields of study beyond cosmology is astrophysics which, to start with, needs good control over spherically symmetric solutions.

Here we come to one of a rather displeasing technical issues of $f(\mathbb T)$ gravity. It is extremely difficult to find exact solutions beyond Minkowski or FLRW spacetimes. In particular, one needs a good understanding of generalised Bianchi identities applied to these models, and even after that, once we have understood that what we have for static spherically symmetric solutions are indeed just two equations for two variables, not three equations, it anyway remains extremely difficult to find exact solutions, even for very simple functions $f$.

In this work we continue our previous investigations \cite{Golovnev:2020las}. In those we show that one can reduce the problem of finding spherically solutions to two equations: the radial equation, and another equation which can be chosen to not depend on the function $f$ at all (except for the most pathological cases). We summarize these findings in Section \ref{sec:ssfT}. We  then present in Section \ref{sec:eqsss} a new approach to looking for solutions which allows to easily obtain some solutions which were previously found using a very heavy machinery of Noether symmetries \cite{Paliathanasis:2014iva,Bahamonde:2019jkf}. 

Arguably, those solutions are not the most physically relevant ones. Consequently, we then apply our approach to representation of asymptotically flat solutions in terms of $1/r$ expansion in Section \ref{sec:as}. The fact that we have an equation independent of $f$ allows us to discuss which solutions are possible at all, for any choice of the function $f$. And then we make conclusions about desired behaviour of the function, if spatially flat solutions are needed. Finally we discuss our findings in Section \ref{sec:ccl}.

\section{Spherical symmetry in $f(\mathbb T)$}
\label{sec:ssfT}

In the notation adopted throughout this work, the dynamical variable is the tetrad field $E^{a}_{\mu}$ instead of the metric itself, which is retrieved from the former by the relation $g_{\mu\nu}=\eta_{ab}E^a_{\mu}E^b_{\nu}$, where the Minkowski metric is adopted as $\eta_{ab}=\text{diag}(1, -1, -1, -1)$. The tetrad lives on top of a teleparallel spacetime, described by the connection
\begin{equation}
\Gamma^{\alpha}_{\mu\nu}=e_a^{\alpha}\left(\partial_{\mu}E^a_{\nu}+\omega^a_{b\mu} E^b_{\nu}\right)    
\end{equation}
where $e$ is inverse to $E$, and $\omega^a_{b\mu}$ is a flat spin connection which will be set to zero in our approach. We define the contortion tensor and the superpotential in terms of the torsion tensor $T^{\alpha}{}_{\mu\nu} = \Gamma^{\alpha}_{\mu\nu} - \Gamma^{\alpha}_{\nu\mu}$ as
$$K_{\alpha\mu\nu}=g_{\alpha\beta}\left(\Gamma^{\beta}_{\mu\nu}-\mathop{\Gamma^{\beta}_{\mu\nu}}\limits^{(0)}\right)=\frac12 \left(T_{\alpha\mu\nu}+T_{\nu\alpha\mu}+T_{\mu\alpha\nu}\right),$$
\begin{equation}
S_{\alpha\mu\nu}=\frac12\left(K_{\mu\alpha\nu}+g_{\alpha\mu}T_{\nu}-g_{\alpha\nu}T_{\mu} \right)
\end{equation}
respectively\footnote{Note that the superpotential is multiplied by a $\frac12$ factor, which differs from many other works}, while the torsion scalar is then simply $\mathbb T=T_{\alpha\mu\nu}S^{\alpha\mu\nu}$ (without a $\frac12$ factor). The quantities related to Levi Civita connection are here and later denoted with $(0)$ superscript.

The action for $f(\mathbb T)$ gravity can be written as
\begin{equation}
S = \dfrac{1}{2\kappa} \int d^4 x\ f(\mathbb T).
\end{equation}

The equations of motion are obtained by varying the previous action with respect to the tetrad field, and are traditionally written in vacuum as
\begin{equation}
2f_{TT}(\mathbb T)  S_{\lambda}{}^{ \mu\nu}\partial_{\mu}\mathbb T - \dfrac{1}{2} \delta^{\nu}_{\lambda} f(\mathbb T) 
 + 2e E^{a}_{\lambda}f_T(\mathbb T) \partial_{\mu}[E e^{\sigma}_a S_{\sigma}{}^{\mu\nu}  ] + 2T^{\rho}{}_{\mu\lambda} S_{\rho}{}^{\mu\nu} f_T(\mathbb T)=0
\end{equation}
but can also be presented in generally covariant form as
$$f_T({\mathbb T})\mathop{G_{\mu\nu}}\limits^{(0)}+2f_{TT}(\mathbb T)S_{\mu\nu\alpha}\partial^{\alpha}{\mathbb T}+\frac12 \left(\vphantom{f^A_B}f(\mathbb T)-f_T(\mathbb T){\mathbb T}\right)g_{\mu\nu}=0.$$

When we work in nonlinear modifications of TEGR, the second derivative $f_{TT}$ is generically non-zero, and we must take care of also the antisymmetric part of equations which are no longer locally Lorentz invariant (in the pure tetrad formalism). It requires to carefully choose the tetrad for a given metric. There is always an option of going for constant $\mathbb T$ solutions which are the same as in GR, however they usually do not respect the symmetry under consideration at the level of tetrads \cite{Nashed:2014sea,Bejarano:2014bca}.

Therefore, the problem of interest is to find a tetrad which simultaneously both respects the symmetry and does not require that $f_{TT}=0$. The working choice for a static spherically symmetric metric which automatically solves the antisymmetric part of the equations of motion is then \cite{Bahamonde:2019jkf}
\begin{equation}
\label{sphshymt}
E^{a}_{\mu} = \left(
\begin{array}{cccc}
A(r) & 0 & 0 & \\
0 & B(r) \sin(\theta)\cos(\phi) & r \cos(\theta) \cos(\phi) & -r\sin(\theta) \sin(\phi) \\
0 & B(r) \sin(\theta) \sin(\phi) & r \cos(\theta) \sin(\phi) & r \sin(\theta) \cos(\phi) \\
0 & B(r) \cos(\theta) & -r\sin(\theta) & 0
\end{array}
\right).
\end{equation}
It corresponds (with the $(+,-,-,-,)$ signature convention) to the torsion scalar
\begin{equation}
\label{torscal}
{\mathbb T} = -\dfrac{2  (B-1)(A-AB+2rA^{\prime}) }{r^2 A B^2}.
\end{equation}
Note that also another tetrad is possible \cite{Ruggiero:2015oka}, different from \eqref{sphshymt} in  the signs of all components with $r$. But they are equivalent by just reverting the signs of $A$ and $B$, since the change of the overall sign of the tetrad brings no physical change at all.

\section{Equations for the spherically symmetric tetrads}
\label{sec:eqsss}

Let us look for spherically symmetric solutions. The equations of motion have a quite complicated form: temporal
\begin{multline}
\label{temporal}
 -\frac12 f - \dfrac{2}{r^2 A B^3}f_{T}\cdot\left(r(B-1)BA' +A(B^2-B+rB')\vphantom{\int} \right) \\
  + \dfrac{8  (B-1)}{r^4 A^2 B^5} f_{TT}\cdot \left(\vphantom{\int}(B-1)\left(A^2(B^2-B-rB')-r^2B A'^2\vphantom{\int}\right) \right. \\
\left. + rA\left(2rA'B' + B\left(A'(1-rB')-rA''\vphantom{A^A_A}\right)+B^2(-A'+rA'') \vphantom{\int}\right) \vphantom{\int}\right)=0,  
\end{multline}
radial
\begin{equation}
 -\frac12 f - \dfrac{2}{r^2 A B^2}f_{T}\cdot \left(A(B-1)+rA'(B-2)\vphantom{\int} \right)=0,
\end{equation}
and angular
\begin{multline}
\label{angular}
-\frac12 f + \dfrac{1}{r^2 A B^3} f_{T}\cdot\left(A\left(B-2B^2+B^3-rB'\vphantom{A^A_A}\right) +r\left(-2B^2A' - rA'B' + B(3A'+rA'')\vphantom{A^A_A}\right) \vphantom{\int}\right) \\
- \dfrac{4(A-AB+rA')}{r^4 A^3 B^5}f_{TT}\cdot\left(
\vphantom{\int}(B-1)\left(A^2(B^2-B-rB')-r^2B A'^2\vphantom{\int}\right) \right. \\
+ \left.rA\left(2rA'B' +B\left(A'(1-rB')-rA''\vphantom{A^A_A}\right) + B^2(-A'+rA'')\vphantom{\int} \right) \vphantom{\int}\right)=0.
\end{multline}
Note that these equations do not depend on possible rescalings of $A$ by a constant factor. The reason is that this is nothing more than a dumb redefinition of the time variable.

One option is to go for extremely pathological cases. If for some constant value of $\mathbb T$ both the function and its first derivative are zero $f(C)=f_T (C)=0$, then if we manage to construct a tetrad with $\mathbb T=C$ equal to this constant, it gives a whole zoo of solutions with effectively switched off gravity. The reason for this is that the $f_{TT}$ terms are multiplied by the gradient of $\mathbb T$, and therefore they vanish too. An explicit example of this pathology with $C=0$ and $f(\mathbb T)=\mathbb T^2$ is given in our paper \cite{Golovnev:2020las}. Using the expression for the torsion scalar (\ref{torscal}), one can see that any tetrad with arbitrary $A$ and either $B=1$ or $B=1+\frac{2rA^{\prime}}{A}$ is a solution for every $f(\mathbb T)$ theory with $f(0)=f_T(0)=0$.

For physically more viable cases with $f_T\neq 0$, we can reduce the system above to two equations which are much simpler. Let us recall that only two of the three equations are independent \cite{Golovnev:2020las}. Therefore, we can conveniently choose to work with only two of them. One of the chosen equations must be the radial one since it only has derivatives of lower order, and fortunately it is also the simplest looking one:
\begin{equation}
\label{radial}
f(\mathbb T) + 4f_T(\mathbb T)\cdot\dfrac{A(B-1) + rA^{\prime}(B-2)}{r^2 A B^2 }=0.
\end{equation}
The temporal \eqref{temporal} and angular (\ref{angular}) equations contain also $f_{TT}$ terms, but with very similar structures there since the most complicated part of them is given by the gradient of $\mathbb T$, which is the same in both. Therefore, we firstly go for a combination of the two equations without the $f_{TT}$ term:
\begin{multline*}
 f(\mathbb T)\cdot \left(A^2 B^2 r^2(B-1) + r^3 A A^{\prime} B^2\vphantom{\int}\right) + f_T(\mathbb T)\cdot \left( 4 A^2B(B-1) - 4A^2 B^2(B-1) + 4rA A^{\prime}\vphantom{\int}\right. \\ 
 \left. - 8r AA^{\prime}B + 4rAA^{\prime}B^2 
  + 4r^2 A^{\prime 2}(B-1)  + 4 r^2A A^{\prime} B^{\prime}  - 4 r^2 A A^{\prime\prime}(B-1) \vphantom{\int}\right) = 0.
\end{multline*}
And once we have obtained this last equation with $f$ and $f_T$ only, we substitute $f$ in favour of $f_T$ by means of the radial equation (\ref{radial}). In the final result, all terms are multiplied by the $f_T$ factor, henceforth it can be cancelled as long as $f_T\neq 0$. It leaves us with an f-independent equation
\begin{equation}
-A^2(B+1)(B-1)^2 + r^2 A^{\prime 2} + r^2 A(A^{\prime} B^{\prime} - A^{\prime \prime}(B-1) )=0.
\label{gencond}
\end{equation}
These two equations (\ref{radial}, \ref{gencond}) above constitute the system of equations which determines all potentially viable solutions, i.e. the ones with $f_T\neq 0$.

\subsection{Some known solutions}

The system of two equations we found gives a much simpler way of searching for solutions. For example, there are some known solutions of $f(\mathbb T)$ gravity of $B=\text{const}$ type which were found by a complicated approach of Noether symmetries \cite{Bahamonde:2019jkf}. We can now find them in a much simpler way. The idea is just to substitute a constant value of $B$ into the f-independent equation and get what should be the function $A$. A possible solution for $A$ then turns out to be a power law in terms of the radius $r$. After that we substitute both $A$ and $B$ into the radial equation which then relates $f$ to $f_T$ and therefore allows us to see for which function $f$ (power of $\mathbb T$) this is indeed a solution. 

To warm up, let us first consider an elementary particular case. We take $B=-1$ and substitute it into the f-independent equation (\ref{gencond}): $r^2({A^{\prime}}^2+2AA^{\prime\prime})=0$, which is then very easy to solve as $A=r^{2/3}$, up to an arbitrary constant factor (another possible solution would be constant $A$ corresponding just to Minkowski metric). It gives ${\mathbb T}=\frac{40}{3r^2}$. And the radial equation (\ref{radial}) then shows that $f_T=\frac{fr^2}{16}=\frac{5f}{6\mathbb T}$ which can easily be solved as $f=\mathbb T^{5/6}$ (up to an overall constant factor again of course).

This is a particularly crazy solution. The spatial geometry is fully Euclidean, and the time rate changes with the radius, apparently not approaching anything flat at infinity. Other known cases are not really more tame, but they also have an angular deficit in the spatial geometry.

Note that we can have even a more general solution above:
\begin{equation}
A=(r-c)^{2/3}
\end{equation}
with arbitrary  constant $c$ which is more general than what has been found by Noether symmetry, though it would correspond to a much more complicated function $f$.

Now turning to the general case of already known solutions \cite{Bahamonde:2019jkf}, suppose $B=b=$ const, an arbitrary constant. Then the f-independent equation (\ref{gencond}) takes the form of
\begin{equation}
-A^2(b+1)(b-1)^2 + r^2 A^{\prime 2} - r^2 A  A^{\prime \prime}(b-1) =0.
\end{equation}
Substituting $A=r^{\alpha}$, we get a solution if
$-(b+1)(b-1)^2+\alpha^2-\alpha(\alpha-1)(b-1)=0$ or 
\begin{equation}
(b-2)\alpha^2-(b-1)\alpha+(b+1)(b-1)^2=0    
\end{equation}
which can be solved as
\begin{equation}
\alpha=\frac{b-1}{2(b-2)}\left(1\pm\sqrt{1-4(b-2)(b+1)}\right).    
\end{equation}
These are real if and only if 
\begin{equation}
\frac12 \left(1-\sqrt{10}\right)\leqslant b \leqslant \frac12 \left(1+\sqrt{10}\right).    
\end{equation}
Then we have for the torsion scalar (\ref{torscal}) 
\begin{equation}
\mathbb T=-\frac{2(b-1)(1-b+2\alpha)}{r^2b^2}  
\end{equation}
which gives for the radial equation (\ref{radial})
\begin{equation}
0=f(\mathbb T) + 4f_T(\mathbb T)\cdot\dfrac{-2 \alpha + b + \alpha b-1}{b^2 r^2}=f(\mathbb T) - 4f_T(\mathbb T)\cdot\dfrac{-2 \alpha + b + \alpha b-1}{2(b-1)(1-b+2\alpha)}\mathbb T    
\end{equation}
which is obviously satisfied in the case of the power-law function 
\begin{equation}
f(\mathbb T)={\mathbb T}^{\frac{(b-1)(1-b+2\alpha)}{2(-2 \alpha + b + \alpha b-1)}}
\label{fTab}
\end{equation}
defining the gravitational theory.

We can check that the particular case from above is indeed reproduced. With $b=-1$ we have two solutions for $\alpha=\frac13 (1\pm 1)$. One is our power $2/3$ which also reproduces the $5/6$ power of $\mathbb T$ in the defining function \eqref{fTab}. Another solution is $\alpha=0$ which corresponds to constant $A$, i.e. Minkowski metric. For non-TEGR cases, this is a non-trivial choice of Minkowski tetrad, the one with negative $B$. However, substituting $b=-1$ and  $\alpha=0$ into the above result for the radial equation, we see that it is never a vacuum solution beyond TEGR since we get the function $f$ with $\mathbb T$ in the power equal to one (the radial equation taking the form of $f-f_T \mathbb T=0$).

\subsection{Further comments on exact solutions}

For any function $f$ when $f_T\neq 0$, every solution of $f(\mathbb T)$ gravity with our choice of tetrad must satisfy the f-independent equation. More to the observation above about non-existence of non-trivial Minkowski solution, note another generic property if $f_T\neq 0$, namely that, by virtue of the f-independent equation (\ref{gencond}), $A=\text{const}$ necessarily requires $B=\pm 1$, while $B=1$ implies $A=\text{const}$.

In a search for exact solutions, one might start from assuming some function $B(r)$ and then solve the second order ODE (\ref{gencond}) for $A$. Another way is to assume some particular $A(r)$ and solve the ODE (\ref{gencond}) for $B$, first order but more non-linear. And it is also possible to start with some relation between $A$ and $B$. After these two functions are found, the radial equation (\ref{radial}) can be solved for the function $f$ since the coefficient in front of $f_T$ is a function of $r$ which can be expressed as a function of $\mathbb T$ (at least locally). 

This procedure can be viewed as reconstructing the function $f$ from the desired properties of solutions \cite{Bamba:2012vg}. Note however that what we have shown is also a generic property (\ref{gencond}), necessary for viable solutions of any $f(\mathbb T)$ theory, and our system of equations (\ref{radial}, \ref{gencond}) looks much simpler for solving than the full initial system, even if a particular function $f$ is chosen from the very beginning.

One idea to use a hypothesis about a possible relation between $A$ and $B$ would be to make the last term in the f-independent equation (\ref{gencond}) vanish. It can be done by $B=1+cA^{\prime}$ with constant $c$ which is arbitrary (obvious a posteriori, or can be found by noticing that the last term is proportional to $\left(\frac{A^{\prime}}{B-1}\right)^{\prime}$). Then the equation requires that either $A=\text{const}$ or a new option with $c^3A^2A^{\prime}+2c^2A^2=r^2$ (a particular solution with $A$ proportional to $r$ is already included in the $B=\text{const}$ cases above).

Yet another interesting thing to note is that the f-independent equation (\ref{gencond}) for $A$, given some arbitrary choice of the function $B(r)$, can be reduced in its full generality to a first order differential equation. Indeed, defining $q(r)=\frac{A^{\prime}}{A}$ which can easily be solved for $A$ if $q$ is found, we find using $\frac{A^{\prime\prime}}{A}=q^{\prime}+q^2$:
\begin{equation}
\label{qB}
-q^{\prime}(B-1)-q^2(B-2)+qB^{\prime}=\frac{(B+1)(B-1)^2}{r^2}
\end{equation}
for the f-independent equation (\ref{gencond}).

This equation looks much simpler for numerical solutions in terms of $q$. However, if we want to find an exact analytical solution, it still seems not so easy. Note however that the first and the last terms in the left hand side can be written together as $-(B-1)^2\left(\frac{q}{B-1}\right)^{\prime}$. Therefore, if we introduce yet another new function $\chi=\frac{q}{B-1}$, the equation (\ref{qB}) turns into
\begin{equation}
-\chi^{\prime}-\chi^2 (B-2)=\frac{B+1}{r^2}.
\end{equation}
In order to separate variables, let us assume that $\frac{B+1}{r^2}=-c(B-2)$ with some constant $c$. It means to choose
\begin{equation}
B(r)=\frac{2cr^2-1}{cr^2+1},
\end{equation}
which asymptotically approaches $2$ at infinity. Depending on the sign of $c$, it has either a zero or a pole at some value of $r$. Let us think that we are looking at sufficiently large $r$, so that there is no such problem (at smaller $r$ it might be changed by presence of matter). The f-independent equation can then be written as
\begin{equation}
\frac{d\chi}{c-\chi^2}=-\frac{3dr}{1+cr^2}
\end{equation}
which can be solved (for positive $c$) as
\begin{equation}
\chi=\sqrt{c}\tanh\left(-3\arctan(\sqrt{c}r) + c_1\right).
\end{equation}
Finally, we see that $q=(B-1)\chi$ and $A=\exp\int q$. At sufficiently large radius, where $\mathbb T$ is a monotonous function of $r$, we might be able to invert this function, $r=r(\mathbb T)$, and solve the radial equation (\ref{radial}) which is guaranteed to be possible since it is then a linear ODE for $f$.
\section{Approximate solutions}
\label{sec:as}

As we have seen above, there are some feasible ways of looking for exact solutions. However, the solutions mostly turn out to be not very simple, and also the methods for making equations analytically solvable do not pay too much attention to desired physical properties.
Therefore it is reasonable to also discuss appropriate approximation schemes. In particular, let us have a look at asymptotically flat solutions.

Some general properties of spherically symmetric solutions can be extracted by performing a perturbative expansion of the functions $A$ and $B$ of the form
\begin{equation}
A(r)=a_0 + \frac{a_1}{r} + \frac{a_2}{r^2} + \ldots,
\label{Aeq}
\end{equation}
\begin{equation}
B(r)=b_0 + \frac{b_1}{r} + \frac{b_2}{r^2} + \ldots.
\label{Beq}
\end{equation}
When putting these expansions into the equation \eqref{gencond}, several branches of solutions can be found, which are about possibilities with some choice of the function $f(\mathbb T)$ in the action.

After that, the radial equation (\ref{radial}) can be used to establish which function $f$ is suitable for getting a chosen solution of the f-independent equation. It should not give any new obstacle, at least as long as $\mathbb T$ is a monotonous function of $r$ therefore allowing for expressing it as $r=r(\mathbb T)$, since the equation is then just a linear first order ODE for the function $f$.

\subsection{Possible solutions (for some choice of the model)}

Substituting the power expansions brings the f-independent equation (\ref{gencond}) to the following form
\begin{equation}
0=-a_0^2(b_0+1)(b_0-1)^2 -\dfrac{a_0 (b_0-1) (2 a_1 b_0^2 + a_0 b_1 (3 b_0 + 1))}{r}+\ldots.
\label{1overr}
\end{equation}
Below we discuss up to orders of up to $1/r^4$ and even higher, but we omit orders higher than $1/r$ in the formula \eqref{1overr}, because in the most general case the formulas get too long and are hard to visualise.

At the zeroth ($1/r^0$) order, the equation \eqref{1overr} has three obvious solutions: $b_0=1$ corresponding to positive $B$ solutions (including the standard Minkowski case), $b_0=-1$ giving negative $B$ solutions which are new classes of solutions, and $a_0=0$ leading to a  metric that is degenerate at infinity, which might be not physically acceptable. Note that the change of the sign of $B$ is an important transformation which does not keep a solution intact, except for the case of TEGR, however the change of the sign of $A$ does not have any effect on our equations, and the reason is that it is just reversal of time.

If we assume $a_0\neq 0$, then by time rescaling this parameter can be set to $a_0=\pm 1$, or simply to $a_0=1$ if we are alright about time reversal. After that, the $b_0=1$ case (as well as $a_0=0$ if we used that) does not entail any new restrictions for the expansion coefficients at the $1/r$ order in \eqref{1overr}. In the case of $b_0=-1$, we get $b_1=a_0 a_1$ or just $b_1= a_1$ if $a_0=1$ is chosen. This might seem a very unusual behaviour (as opposed to $b_1=-a_1$), however since $B$ and $A$ are of opposite signs in this case, this is just the usual property of the Schwarzschild solution of TEGR, while the choice of $a_0=-1$ with $b_0=-1$ also includes the Schwarzschild case but with the choice of negative signs for both $A$ and $B$.

Below we illustrate bifurcations of possible spherically symmetric asymptotically flat solutions of $f(\mathbb T)$ with the choice of $a_0=1$.

\textbf{Case X.} In the most conservative case of $b_0=1$ we have at $1/r^2$ order
\begin{equation}
 a_1^2 - a_0 a_1 b_1 - 2 a_0^2 b_1^2=0
\end{equation}
which has two possible solutions (which we give for $a_0=1$ choice), either $a_1=-b_1$ or $a_1=2b_1$. 

\textbf{Case XX.} When following the first branch $a_1=-b_1$, the term going with $1/r^3$ is
\begin{equation}
\dfrac{2 b_1 (-4 a_2 + b_1^2 - 2 b_2)}{r^3}
\end{equation}
which has two solutions $b_1=0$ and $b_2=-2a_2+\frac12 b_1^2$. The $1/r^4$ term is
\begin{equation}
\dfrac{4 a_2^2 - 15 a_3 b_1 + a_2 b_1^2 - 2 a_2 b_2 + 5 b_1^2 b_2 - 2 b_2^2 - 
 5 b_1 b_3}{r^4}
 \label{rfourth}
\end{equation}
where we have imposed all previous restrictions, except choosing an option from the last one.

\textbf{Case XXX.} In the case of $b_1=0$ we get \eqref{rfourth} as $4a_2^2-2a_2b_2-2b_2^2=0$ which gives either $b_2=a_2$ or $b_2=-2a_2$. 

\textbf{Case XXY.} In another case of $b_2=-2a_2+\frac12 b_1^2$ we get \eqref{rfourth} as $b_1(2b_1^3-6b_1a_2-5b_3-15a_3)=0$ which gives either $b_1=0$ now or $b_3=-3a_3+\frac25 b_1^3-\frac65 b_1a_2$.

\textbf{Case XY.} The second branch $a_1=2b_1$ (and $a_0=b_0=1$) has the coefficient of the $1/r^3$ order as
\begin{equation}
\dfrac{b_1 (4 a_2 - 13 b_1^2 - 4 b_2)}{r^3}.
\end{equation}
We can vanish this term by either doing $b_1=0$ again, or by choosing $b_2=a_2-\frac{13 b_1^2}{4}$.

The \textbf{Case XYX} of $b_1=0$ is then the same as the \textbf{Case XXX} above since the choice of $b_1=0$ entails $a_1=0$ in both cases.

\textbf{Case Y.} What happens in the $b_0=-1$ case, can also be easily seen. After taking $a_0=1$ and $a_1=b_1$, we get $(12 a_2 - 4 b_2)/r^2$, which establishes $b_2=3a_2$. Next term is 
\begin{equation}
   \dfrac{ (2 a_1^3 + 8 a_1 a_2 + 24 a_3 - 4 b_3)}{r^3}
\end{equation}
which sets $b_3$ as
$b_3 = 6a_3+2a_1a_2+\frac12 a_1^3$.

The first steps of bifurcations of possible solutions are presented in the Figure \ref{SphSymmTikz}. We start with the choice of $a_0=1$ since it is always possible by a proper choice of time variable (in every good physical case of $a_0\neq 0$). After that every row of the diagram shows possible solutions at the increasing orders of $1/r$.

\tikzstyle{ba} = [rectangle, draw,
text width=3em, text centered, rounded corners, minimum height=2em]
\tikzstyle{bg} = [rectangle, draw,
text width=4em, text centered, rounded corners, minimum height=2em]
\tikzstyle{bh} = [rectangle, draw,
text width=5em, text centered, rounded corners, minimum height=2em]
\tikzstyle{tbg} = [rectangle, draw,
text width=7em, text centered, rounded corners, minimum height=2em]
\tikzstyle{line} = [draw, -latex']

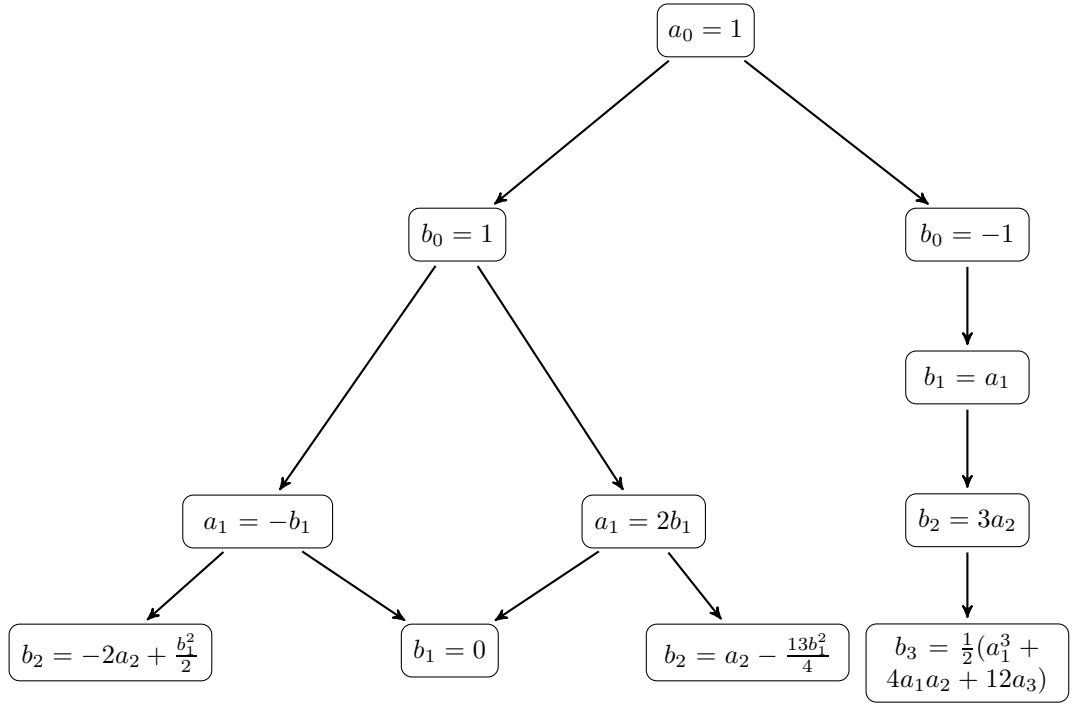
\begin{figure}[htbp!]
\begin{center}
\begin{tikzpicture}[node distance = 1.9cm, auto]
	\node [ba] (r00) { $a_0=1$ };
	\node [ba, below left=2cm and 2cm of r00] (r01) {$b_0=1$};
	\node [bg, below right=2cm and 2cm of r00] (r02) {$b_0=-1$};
	\node [bg, below of=r02] (r10) {$b_1=a_1$};
	\node [bh, below left=3.1cm and 1cm of r01] (r21) {$a_1=-b_1$};
	\node [bg, below right=3.1cm and 1cm of r01] (r22) {$a_1= 2b_1$};
	\node [bg, below of=r10] (r23) {$b_2=3a_2$}; 
	\node [tbg, below left=1cm and -0.4cm of r21] (r32) {$b_2=-2a_2+\frac{b_1^2}{2}$};
	\node [ba, below left=1cm and 1.1cm of r22] (r33) {$b_1=0$};
	\node [tbg, below right=1cm and -0.8cm of r22] (r34) {$b_2=a_2-\frac{13b_1^2}{4}$};
	\node [tbg, below of=r23] (r35) {$b_3=\frac12(a_1^3+4a_1 a_2 + 12a_3)$};
	\draw[pil,->] (r00) -- node  {{}} (r01);
	\draw[pil,->] (r00) -- node  {{}} (r02);
	\draw[pil,->] (r01) -- node  {{}} (r21);
	\draw[pil,->] (r01) -- node  {{}} (r22);
	\draw[pil,->] (r02) -- node  {{}} (r10);
	\draw[pil,->] (r10) -- node  {{}} (r23);
	\draw[pil,->] (r23) -- node  {{}} (r35);
	\draw[pil,->] (r21) -- node  {{}} (r32);
	\draw[pil,->] (r21) -- node  {{}} (r33);
	\draw[pil,->] (r22) -- node  {{}} (r33);
	\draw[pil,->] (r22) -- node  {{}} (r34);
	\end{tikzpicture}
\end{center}
\caption{Bifurcation of possible spherically symmetric asymptotically flat solutions of $f(\mathbb T)$ gravity. The constants $a_i$ and $b_i$ are defined as in Eqs.\eqref{Aeq} and \eqref{Beq}. } 	\label{SphSymmTikz}
\end{figure}

One good check is that Schwarzschild is a solution of the f-independent equation (\ref{gencond}). It can be established explicitly and exactly, but it is also nice to see how it has its place in our approximation procedures. We take
\begin{equation}
A=\sqrt{1-\frac{2M}{r}}=1-\frac{M}{r}-\frac{M^2}{2r^2}-\frac{M^3}{2r^3}+\ldots ,
\end{equation}
\begin{equation}
B=\frac{1}{\sqrt{1-\frac{2M}{r}}}=1+\frac{M}{r}+\frac{3M^2}{2r^2}+\frac{5M^3}{2r^3}+\ldots .
\end{equation}
and it is obviously the case of $a_0=b_0=1, \quad b_1=1=-a_1, \quad b_2=-2a_2+\frac12 b_1^2, \quad b_3=-3a_3+\frac25 b_1^3-\frac65 b_1a_2$. Another option to construct the Schwarzschild metric is to change the sign of $B$ keeping $A$ the same. It corresponds to $a_0=1=-b_0, \quad b_2=3a_2, \quad b_3 = 6a_3+2a_1a_2+\frac12 a_1^3.$ And in both cases there is a possibility of taking all coefficients beyond the $0$-th order vanishing. It is not surprising since Minkowski space is also a possible solution.

For completeness, let us mention that the unphysical case of $a_0=0$ would lead at the $\frac{1}{r^2}$ order to a cubic equation $-b_0^3+b_0^2-b_0+2=0$ which has only one real root, but rather complicated and unpleasant. We do not consider this case at any more detail.

\subsection{Finding a proper function $f$}

After having found some possible (in terms of f-independent equation) solution, one can go on to find a function $f(\mathbb T)$ which would make such solution possible. Since we have a feasible solution at hand, then we know the function $\mathbb T(r)$ explicitly, therefore if we are able to solve for the radial coordinate as $r=r(\mathbb T)$, then the linear ODE for $f$ must be easily solvable.

For example, if we choose the case of $a_0=1$, $b_1=a_1=0$ and $b_2=-2 a_2$  from the previous subsection, we have $A=1+\frac{a}{r^2}$ and $B=1-\frac{2a}{r^2}$ with the leading order at infinity in the torsion scalar as $\mathbb T\approx-\frac{8a^2}{r^6}$. 

The leading order in $\mathbb T$ would be enough for our purposes in the illustration to follow. However let us first see how it works at higher orders. Let us take a more precise  solution as
\begin{equation}
A=1+\frac{a}{r^2}+\frac{a_4}{r^4}+\mathcal O\left(\frac{1}{r^6}\right),
\end{equation}
\begin{equation}
B=1-\frac{2a}{r^2}+\frac{b_4}{r^4}+\mathcal O\left(\frac{1}{r^6}\right).
\end{equation}
Substituting it into the f-independent equation (\ref{gencond}), we see that all orders up to $1/r^5$ are satisfied while at the $1/r^6$-th one has the coefficient of $-4a^3+10ab_4+40aa_4$ which requires either $a=0$ (which we do not want) or $b_4=-4a_4+\frac25 a^2$. The $1/r^6$ terms in $A$ and $B$ would be restricted by the $1/r^8$ order in the equation (\ref{gencond}).

The torsion scalar (\ref{torscal}) can then be found as 
\begin{equation}
\mathbb T=-\frac{8a^2}{r^6} -\frac{16a^3+32 aa_4}{r^8}+\mathcal O \left(\frac{1}{r^{10}}\right).
\end{equation}
In the leading order it can be solved for $r$ as 
\begin{equation}
\label{rT}
r^2=\frac{2a^{2/3}}{(-\mathbb T)^{1/3}}.
\end{equation}
If we want, we can find it with more precision choosing for simplicity $a_4=0$ (and therefore $b_4=\frac25 a^2$):
\begin{equation}
r^2=\left(\frac{8a^2}{-\mathbb T}\left(1+\frac{2a}{r^2}\right)\right)^{1/3}\approx \left(\frac{8a^2}{-\mathbb T}\left(1+a^{1/3}(-\mathbb T)^{1/3}\right)\right)^{1/3}\approx \frac{2a^{2/3}}{(-\mathbb T)^{1/3}}+\frac{2a}{3}.
\end{equation}

The coefficient in the radial equation (\ref{radial}) can be found in the leading order as
\begin{equation}
\dfrac{4[A(B-1) + rA^{\prime}(B-2)]}{r^2 A B^2}=\frac{8a^2+4b_4+16a_4}{r^6}+{\mathcal O}\left(\frac{1}{r^8}\right)=-\left(1+\frac{b_4}{2a^2}+\frac{2a_4}{a^2}\right){\mathbb T} +{\mathcal O}\left({\mathbb T}^{4/3}\right).
\end{equation}
Given the necessary condition of $b_4=-4a_4+\frac25 a^2$, the radial equation (\ref{radial}) takes the form of
\begin{equation}
f= \frac65 f_T\left( \mathbb T+{\mathcal O}\left({\mathbb T}^{4/3}\right)\right)
\end{equation}
which implies
\begin{equation}
f(\mathbb T)={\mathbb T}^{5/6} +{\mathcal O}\left({\mathbb T}^{7/6}\right)
\end{equation}
as the leading order behaviour in the small values of $\mathbb T$.

One can go further and find corrections to the function $f$ by using higher order terms in the definition of functions $A$ and $B$, and also using the corrections to the expression of $\mathbb T$ in terms of $r$.

\section{Conclusions}
\label{sec:ccl}

We have presented a systematic approach to the search of static spherically symmetric solutions in $f(\mathbb T)$ gravity. For physically viable solutions, i.e. those for which the gravity is not effectively switched off by vanishing of $f_T$, the system of equations is reduced to two simple ones. One of them is f-independent and describes the properties which are necessary for the functions $A,B$ describing these solutions, with any choice of the function $f$. The second one is the radial component of the equations of motion, and it can be viewed as an equation (first order linear ODE) which determines the function $f$ needed for reproducing a chosen possible solution. 

Our approach allows to easily find some known exact solutions, which were obtained through a more intricate method of Noether symmetries. Later, we have applied our approach to the most general asymptotically flat solution expanded in terms of $1/r$, and discuss which solutions are plausible for any choice of the function $f$. Finally, some conclusions on the desired behaviour of the function $f$ are possible to be drawn for certain branches.

We expect this investigation will help the teleparallel community to study spherically symmetric solutions and their applications in astrophysics \cite{Bahamonde:2019zea,Bahamonde:2020bbc}, since it is of rapidly increasing importance to confront the theory with observations beyond simple cosmological regimes and/or gravitational waves.

\section*{Acknowledgments}
The authors thank A. Awad and S. Bahamonde for helpful discussions. M.J.G. was funded by FONDECYT-ANID postdoctoral grant 3190531.

\end{document}